# Ballistic conductance in oxidized Si nanowires

*Giorgos Fagas[*] and James C. Greer*

Tyndall National Institute, Lee Maltings, Prospect Row, Cork, Ireland

georgios.fagas@tyndall.ie

---

[*] Author to whom correspondence should be addressed: georgios.fagas@tyndall.ie

Abstract

The influence of local oxidation in silicon nanowires on hole transport, and hence the effect of varying the oxidation state of silicon atoms at the wire surface, is studied using density functional theory in conjunction with a Green's function scattering method. For silicon nanowires with growth direction along [110] and diameters of a few nanometers, it is found that the introduction of oxygen bridging and back bonds does not significantly degrade hole transport for voltages up to several hundred millivolts relative to the valence band edge. As a result, the mean free paths are comparable to or longer than the wire lengths envisioned for transistor and other nanoelectronics applications. Transport along [100]-oriented nanowires is less favorable, thus providing an advantage in terms of hole mobilities for [110] nanowire orientations, as preferentially produced in some growth methods.

Semiconductor nanowires are candidates for nanoelectronics applications and recent progress in growth and fabrication techniques has allowed both bottom-up [1-4] and top-down [5-7] demonstrations of devices with diameters of only a few nanometers. Silicon nanowires (SiNWs), in particular, offer a path to the ultimate limits of scaling with the potential advantage of ease of integration with existing materials in use in semiconductor technologies. An important factor for the use of silicon nanowires is the unavoidable formation of oxygen-derived defects at the surface of a pure semiconductor wire core due to oxidation. Surface oxidation of group IV semiconductor nanowires is a naturally occurring process which results in an oxide sheath around the wire core [8, 9]. Although the native oxide can be removed via HF-etching to provide hydrogenated surfaces that are, for example, subsequently functionalized for sensor applications [10], a residual density of oxygen impurities is expected. In several applications, further treatment of the oxide is instrumental to device operation, as for the case of defect states near the core/oxide interface believed to be responsible for part of the nanowire photoluminescence spectrum [11, 12]. A noticeable experimental demonstration of the impact of the oxide properties on conductance is given in Ref 4, where it was demonstrated that two orders of magnitude enhancement of the hole mobility can be achieved after chemical modification of the $SiO_x$ surface.

In general, oxygen impurities promote the silicon atoms at the outer part of the core to various oxidation states and distort the underlying lattice periodicity, thus, forming scattering centers for charge carriers. To understand the impact of oxygen related defects on electrical conduction, it is instructive to view the defects as a source of roughness on the atomic scale. It is well-known that surface roughness is a key factor for charge carrier transport as it reduces mobility in planar metal-oxide-semiconductor field effect transistors (MOSFETs) [13]. More recently, the detrimental effect of surface roughness was shown to be even more pronounced as the body of the silicon is scaled down to a few nanometers in MOSFETs made of films of ultrathin silicon-on-insulator [14]. By comparison, SiNWs are an extreme realization of a thin semiconductor body with confinement in an additional spatial dimension yielding one-dimensional transport. Hence, scattering from defects becomes unavoidable and an assessment of the impurity-limited carrier transport becomes necessary.

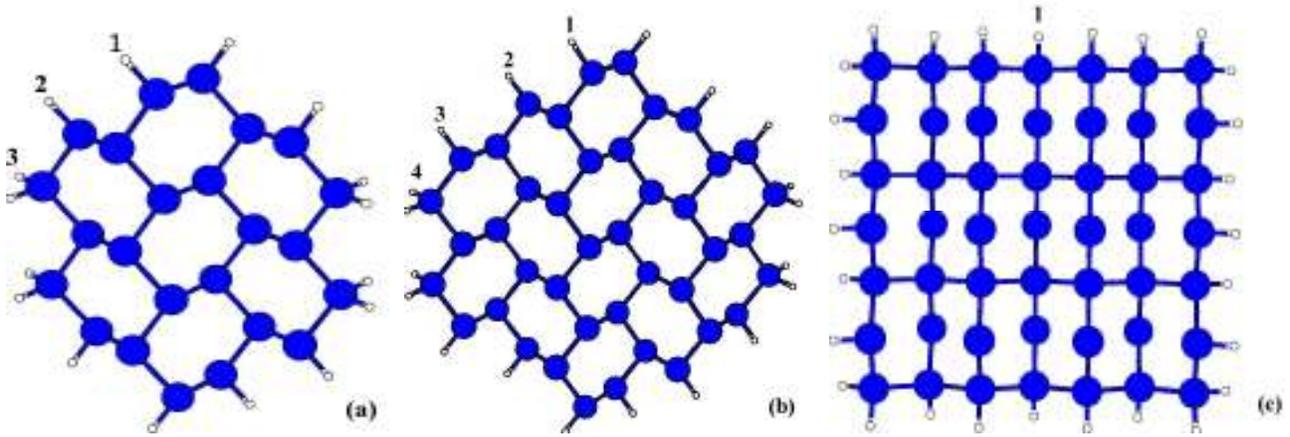

Figure 1: The cross-sections of [110]-oriented hydrogenated SiNWs with diameter W = 1.15nm and 1.54nm, respectively, are seen from a slight angle from the wire axis in (a) and (b). The cross-section of an [100]-oriented wire with W= 1.63nm is shown in (c). W measures the distance between furthest Si atoms in cross-sectional slice of the wire.

In this letter, we investigate the transport properties of ultrathin SiNWs with surface oxidized silicon atoms by means of atomic-scale models (Figures 1 and 2). Such wires have been synthesized by Ma et al [15] and alongside carbon nanotubes are targeted as the basic elemental building blocks for ultimate nanoelectronics. Previous studies on the conducting properties of oxidized SiNWs have mostly relied on a continuum model of oxide roughness [14, 16, 17] and focused on wire diameters larger than 3 nm. Such approaches give an invaluable insight into the physical mechanisms of electron scattering in these nanostructures. But these models are not straightforward to apply on the scale of a few nanometer wire cross sections [18] and input from a microscopic description is required for an accurate picture of electron scattering from the surface oxide. For example, resonant backscattering from defect states cannot be accounted for without a microscopic scattering description; as well, the precise strength and spatial extent of the scattering-center potential must also be determined from atomic-scale models. A first study of an atomistic treatment of oxide interface roughness was taken in Ref. 19, where a semi-empirical model was employed to investigate the impact of long-range thickness variation in a 2.1x2.1 nm$^2$ silicon channel to the device characteristics of triple-gate SiNW transistors. The surface scattering in this case arises from geometric roughness. For our purposes, Density Functional Theory (DFT) calculations are applied on the atomic scale. This enables an accurate treatment of the different oxidation states of the surface silicon atoms and the dependence on the optimized atomic positions which are difficult to describe for semiconductor/oxide interfaces using semi-empirical methods [20].

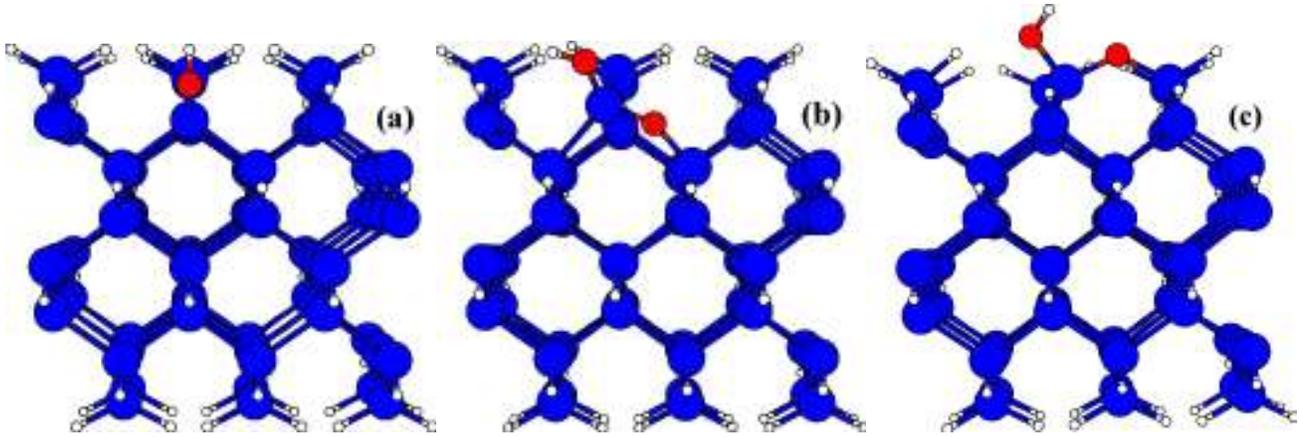

Figure 2: Examples of an (a) Si-O-H bond, (b) Si-O-Si back-bond and (c) Si-O-Si bridge-bond.

To isolate the main effect of defects in an oxidized surface, namely, the differences in the Si potential due to varying oxidation, we employ a simplified scheme. We take hydrogenated reference structures with cross-sections as shown in Figure 1 and start from the neutral oxidation state Si$^{0+}$ at the surface of the hydrogenated wires. We then consider promotion to the Si$^{1+}$ and Si$^{2+}$ states via introducing oxygen at the wire surface as shown in Figure 2. The resulting Si-O-H and Si-O-Si bonds are the simplest cases of oxygen bridge- and back- bonds [18, 21]. This chemically motivated scheme not only offers a practical advantage when compared to nanowires wrapped by an oxide but is also of relevance to sensor applications [10] and porous silicon [11, 12]. Our approach captures the effects of carrier scattering from Si$^+$ atomic sites at the surface of small diameter nanowires. Our findings will indicate that there are certain bias windows for which the actual oxidation states of the surface silicon atoms do not significantly alter the transport properties of a nanowire. However, the treatment of fully oxidized nanowire surfaces may be needed in specific applications, e.g., in predicting the device characteristics of a MOSFET structure.

In what follows, we focus the discussion on nanowires with [110] and [100] orientations. SiNWs as synthesized in Ref 3 are oriented predominantly along the [110] direction. The [100] wire axis is most relevant in the model of porous silicon as intercalated SiNWs [11, 12] and is common in top-down fabricated FETs [5]. We demonstrate that the oxygen impurities depicted in Figure 2 are transparent for hole transport over a wide energy window for the lowest energy subbands, i.e., when transport is quasi-one dimensional. For typical defect densities, this implies quasi-ballistic transport as the mean free path can be quite large compared to the device length. This result could not be determined without a first-principles calculation that reveals the small short-range perturbation introduced by the defects. We find the electronic spectrum of the oxidation states is responsible for typically narrow resonant backscattering features. However, significant resonant backscattering may occur at higher subband energies from back-bond defects at specific sites. In general, [110] SiNWs exhibit higher transparency than the [100]-oriented wires due to a combination of their lower effective mass and the electronic density of states.

The theoretical framework of calculating transport coefficients has been frequently reviewed [22]. Within the context of defect scattering in SiNWs, it has been applied to study the effect on carrier transport of dopant impurities [23, 24] and surface chemical functionalization [25]. Other reports include transport studies of the effect of the SiNWs orientation and diameter variation [26-28] as well as the influence of the metal contacts [29]. The first step comprises expressing the Hamiltonian in a localized basis set. The second step requires the standard calculation of the total channel transmission from the quantum mechanical scattering matrix using a Green's function technique [22]. In the Landauer formalism, the transmission function expresses the low-bias low-temperature conductance in units of $2e^2/h$ and can be used to extract the mean free path $l_{imp}$ for impurity scattering of charge carriers injected at energy E [30]. Below we estimate $l_{imp}(E)$ simply from the conductance of single-impurities as described in Ref 23.

For the work presented here, our in-house transport simulator TIMES [31] is combined with a DFT-approach based on the Harris-Foulkes functional [32]. The latter expresses the energy around a reference atomic ground-state energy which allows building an efficient tight-binding (DFTB) approximation with a minimal number of input parameters [33]. The wavefunction for valence electrons is expanded in Slater-type orbitals yielding a Hamiltonian structure similar to an extended Hückel theory. The DFTB method has been recently used to investigate the properties of silicon nanomaterials, e.g., the electronic structure of SiNWs with different surface passivations [34] and cross-sections [35], as well as the photoluminescence of silicon quantum dots [36].

A tetragonal supercell extending five lattice constants along the wire and 100 nm perpendicular to the wire is used in the calculations. This suffices to converge our results as the DFTB interatomic interactions and overlap integrals in the considered SiNWs typically vanish for distances larger than 2.5 times the size of covalent bonds [37]; the lattice constant is 3.84 Å and 5.43 Å in the-[110] and [100]-orientation, respectively. A full relaxation of the input atomic geometries is performed such that the force on each atom is less than 0.015eV/Å. Predicted bond lengths are in good agreement with typical experimental values. The Si-Si, Si-H, Si-O, and O-H bond lengths are found to be 2.3Å, 1.5Å, 1.7Å and 0.98Å, respectively. Interestingly, the lattice distortion does not extend significantly beyond the length of a silicon unit cell around the oxygen defects. Charge population analysis on the Si species of the small-diameter [110]-oriented SiNW yields 0.35 to 0.48 (0.74 to 0.78) deficit electron charge for singly (doubly) oxygen-bonded atoms, respectively, in line with expectations from the formal Si oxidation states. Experimentally p-doping is preferred [4] and our parameterization is most accurate for valence band states due to the minimal basis used, hence, we have focused on hole transport [37, 38]. Also, a significant contribution from inter-subband scattering [16] should be most prevalent in valence bands as these states are closely spaced in energy. Nevertheless, based on spectral properties from less approximate plane-wave DFT schemes [39], we anticipate similar results for the lowest electron channels in terms of observing ballistic transport in these systems.

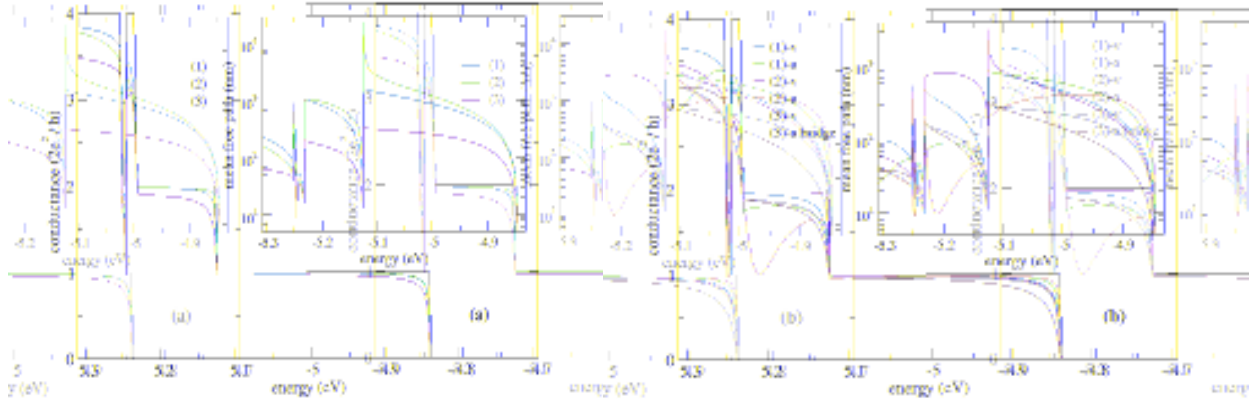

Figure 3: Transport properties of the [110]-oriented SiNW with diameter W = 1.15nm. The conductance is shown for (a) Si-O-H and (b) Si-O-Si bonds at the various sites of Figure 1(a) as indicated. The notation 'a' ('v') in (b) indicates back-bonds oriented along (vertical to) the wire axis. The black lines in both panels indicate the ideal conductance of the oxygen-free hydrogenated SiNW. The extracted mean free path for a defect density $n = 5 \cdot 10^{19} cm^{-3}$ (mean distance between impurities d = 9.6nm) is plotted in the insets.

In Figure 3, we summarize the results for p-type transport in the small-diameter [110]-oriented SiNW with oxygen impurities at non-symmetric sites. Both the Landauer conductance and the mean free path are shown. For comparison, the conductance $G_c$ of the oxygen-free hydrogenated SiNWs is also plotted. It is evident from Figure 3(a) that a single hydroxyl species is almost transparent to the injection of holes as the conductance remains close to its ideal value. The variation of the Si-Si bond lengths around the OH-impurity site is of the order of 0.01Å, implying that the scattering potential created by the Si oxidation is weak. The conductance shown in Figure 3(b) also indicates the same effect; here, silicon sites are promoted to $Si^{2+}$ by introduction of an additional back-bond (bridge-bond) oxygen Si-O-Si pointing the wire core (on the surface). No large variation is observed beyond the expected slightly stronger scattering due to the additional local lattice distortion. The small effect of the oxidation-induced potential is further supported by control simulations (not shown) where we reduced the level of oxidation to $Si^{1+}$ by hydrogenating the Si-O-H bond and obtained similar conductance spectra to Figure 3(b). A prominent exception is the large dip in the conductance line marked as (2)-a in Figure 3(b) around 0.35 eV below the valence band maximum. This feature of resonant backscattering is related to quasi-bound oxygen states and will be discussed below.

The impurity-induced mean free path $l_{imp}(E)$ is calculated as follows [23]. In the diffusive regime, the conductance is $G = G_c / (1+L/l_{imp})$, where $1/G_c$ is the contact resistance of a nanowire with ideal conductance $G_c$. Considering that single-impurities will classically add up as resistances in series, the conductance is also given by $G = G_c / (1+ G_c/G_s \cdot L/d)$. Here, $1/G_s$ is the resistance of a single-defect and is extracted by taking the inverse of the conductance given in Figure 3 and subtracting $1/G_c$; d is the mean distance between impurities. Equating the two expressions for the conductance yields $l_{imp} = (G_s/G_c) \cdot d$. Results for the typical oxygen-defect density of $n = 5 \cdot 10^{19} cm^{-3}$ are shown in the insets of Figure 3. Notably, the mean free path is predominantly in excess of 100 nm for much of the spectrum. This implies that for typical device lengths the majority of the injected carriers do not experience a scattering event, thereby, yielding ballistic transport for these carriers. It is interesting that scattering from other surface impurities is also associated with large $l_{imp}$. For example, it has been found [23, 40] that surface dopants have a smaller scattering cross-section than dopants located at in nanowire core.

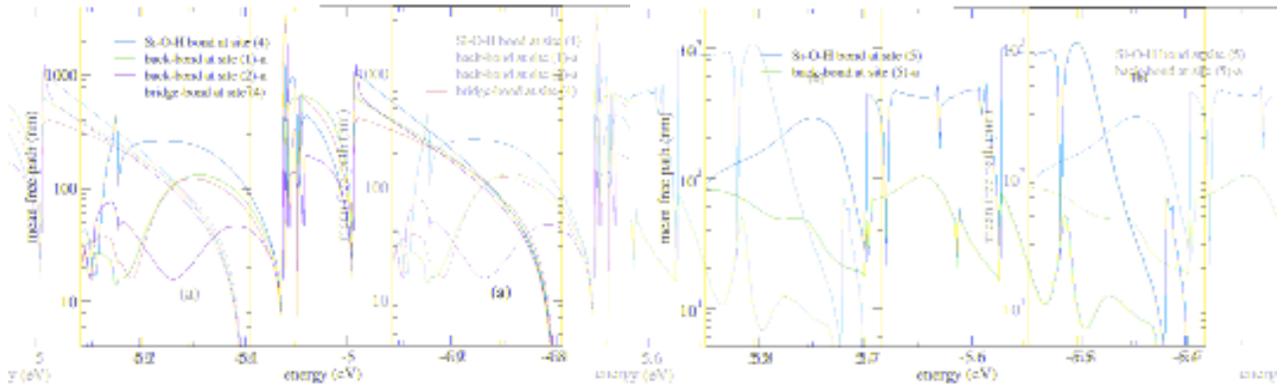

Figure 4: Mean free path for SiNWs oriented along the (a) [110]- and (b) [100]-direction with diameter W equal to 1.54nm (Fig 1(b)) and 1.63mn (Figure 1(c)), respectively. The defect density is n = 5·10$^{19}$cm$^{-3}$ with mean distance between impurities d = 6.14nm in (a) and 7.1nm in (b).

The diameter and growth-orientation dependence of the mean free path is demonstrated in Figure 4. It is well known that quantum confinement causes a diameter-dependent blue-shift in the band gap of nanowires as the diameter decreases [11, 12, 29, 34, 41]; at fixed diameter, the shift is smaller in SiNWs grown along the [110]-direction than along [100] [29]. Linked to the confinement effect is the red-shift of the subbands of the [110]-oriented SiNW in Figure 4(a) (compared to Figure 3). As the spectral features of the oxygen defect states do not depend strongly on increasing the diameter, $l_{imp}$ clearly retains its high value for the lowest subbands and transport is essentially ballistic. However, a crystallographic change in a nanowire's orientation has a considerable effect. Scattering is more pronounced in a [100]-oriented SiNW with similar diameter, thus, resulting in smaller mean free paths (Figure 4(b)). The origin of this effect is twofold. First, the effective mass along the [100]-direction is larger. As slower carriers scatter more, $l_{imp}$ decreases (in fact, in a simple effective-mass theory model $l_{imp}$ is inversely proportional to the mass [40]). This agrees with the conclusion of Ref 19 that transport along the [110]-direction is less sensitive to long-range surface roughness compared to the [100]-orientation. But for the atomic-scale roughness considered here, the second and most important contribution to scattering is that more oxygen-defect states lie within the energy window spanned by the valence-band states of the [100]-SiNW.

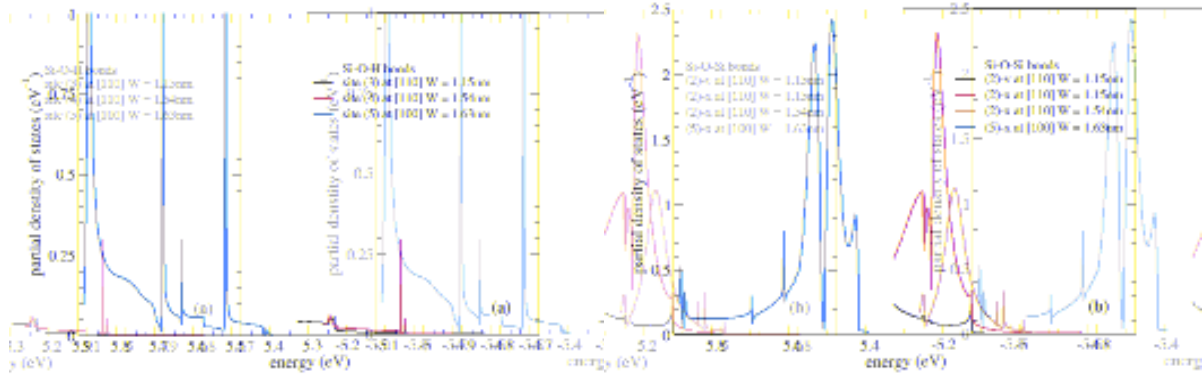

Fig 5: Density of states at the oxygen site for (a) Si-O-H and (b) Si-O-Si bonds. The site notation is as in Fig 3.

To explain the observed features in the transport spectra, in Figure 5 we plot the density of states (DOS) as extracted from the imaginary part of the Green's function at the oxygen site in Si-O-H and Si-O-Si bonds. We first note the large contribution of oxygen states over the whole energy window in the case of [100]-oriented SiNW, which is in agreement with the enhanced scattering. There is also a clear correlation between DOS-spikes and dips in the mean free path shown in Figure 4(b); other dips are associated with the contribution of additional subbands as the energy is varied and the fact that their corresponding group velocity vanishes at the entry point. DOS-related dips in the transport properties are caused by resonant backscattering as explained within the theory of Fano (anti-)resonances in quantum transport [42], which implies the destructive interference between the SiNW transport channels and weakly-coupled oxygen states. For the [110]-grown SiNWs, apart from very narrow features no appreciable density of states is found on oxygen for most defect configurations. The only significant contribution occurs for the Si-O-Si bond at the configuration marked as (2)-a resulting in an anti-resonance with a width of the order of room temperature.

In conclusion, using methods based on first-principles electronic structure and scattering theory, we find that the effect of local variations in surface Si oxidation number does not give rise to large scattering cross sections for hole transport at low bias voltages in [110]-oriented silicon nanowires. Hence, mobility degradation cannot be associated with the electronic states and lattice distortions arising from surface oxidation in nanowires with cross sections of a few nanometers and lengths shorter than the resulting scattering lengths (i.e. on the order of 100 nm and below). This result is encouraging, as the preferred growth direction of chemically synthesized nanowires is [110] and the local oxidation, even for surface treated wires, is unavoidable. On the other hand, Si oxidation-state variations may result in higher carrier scattering for transport along a [100]-oriented silicon nanowire and this should also be considered when selecting nanowire orientations within top-down nanowire fabrication.

**Acknowledgements**

We are grateful to the Science Foundation Ireland for supporting this work. We would also wish to acknowledge the SFI/HEA Irish Centre for High-End Computing for the provision of computational facilities.